\documentclass[
 reprint,
 amsmath,amssymb,
 aps,
]{revtex4-2}

\usepackage{graphicx}
\usepackage{dcolumn}
\usepackage{bm}
\usepackage{xcolor}
\usepackage{hyperref}
\usepackage[normalem]{ulem}

\hypersetup{
colorlinks=true,
citecolor=blue,
linkcolor=blue,
urlcolor=blue}

\begin{document}

\preprint{APS/123-QED}

\title{Impact of late-time neutrino emission on the diffuse supernova neutrino background}

\author{Nick Ekanger$^1$}
\email{enick1@vt.edu}
\author{Shunsaku Horiuchi$^{1,2}$}
\author{Kei Kotake$^{3,4}$}
\author{Kohsuke Sumiyoshi$^5$}
\affiliation{${}^1$Center for Neutrino Physics, Department of Physics, Virginia Tech, Blacksburg, VA 24061, USA.}
\affiliation{${}^2$Kavli IPMU (WPI), UTIAS, The University of Tokyo, Kashiwa, Chiba 277-8583, Japan}
\affiliation{${}^3$Department of Applied Physics, Fukuoka University, Nanakuma Jonan 8-19-1, Fukuoka 814-0180, Japan}
\affiliation{${}^4$Research Institute of Stellar Explosive Phenomena, Fukuoka University, Nanakuma Jonan 8-19-1, Fukuoka 814-0180, Japan}
\affiliation{$^5$National Institute of Technology, Numazu College of Technology, Ooka 3600, Numazu 410-8501, Japan}

\date{\today}

\begin{abstract}
In the absence of high-statistics supernova neutrino measurements, estimates of the diffuse supernova neutrino background (DSNB) hinge on the precision of simulations of core-collapse supernovae. Understanding the cooling phase of protoneutron star (PNS) evolution ($\gtrsim1\,{\rm s}$ after core bounce) is crucial, since approximately 50\% of the energy liberated by neutrinos is emitted during the cooling phase. We model the cooling phase with a hybrid method by combining the neutrino emission predicted by 3D hydrodynamic simulations with several cooling-phase estimates, including a novel two-parameter correlation depending on the final baryonic PNS mass and the time of shock revival. We find that the predicted DSNB event rate at Super-Kamiokande can vary by a factor of $\sim2$--3 depending on the cooling-phase treatment. We also find that except for one cooling estimate, the range in predicted DSNB events is largely driven by the uncertainty in the neutrino mean energy. With a good understanding of the late-time neutrino emission, more precise DSNB estimates can be made for the next generation of DSNB searches.
\end{abstract}


\maketitle

\section{\label{sec:intro}Introduction}

When the cores of massive ($\sim8\,M_{\odot}$) stars collapse, $\gtrsim10^{53}\,{\rm erg}$ of gravitational binding energy is released via neutrinos \cite{Kotake:2005zn,2016NCimR..39....1M,2017janka,2021burrows}. In 1987, one such nearby collapse allowed the observation of tens of neutrino events \cite{Hirata:1988iyh,Bionta:1987qt,ALEXEYEV1988209}. Since the historic 1987 supernova, the astrophysical community awaits the next nearby core collapse which will provide a high-statistics neutrino event sample and a wealth of information about, e.g., the dynamical stellar properties and the core-collapse explosion mechanism (see e.g., Refs.~\cite{2012scholberg,2016NCimR..39....1M,Horiuchi:2018ofe} and motivated by simulations, e.g.,  Refs.~\cite{2019suwa,2021li,2022nakazato}).

The core-collapse supernovae (CCSNe) that have occurred over cosmological history give rise to a diffuse background of neutrinos (i.e., the diffuse supernova neutrino background or ``DSNB'', see \cite{2004ando,2010beacom,2016lunardini,2020vitigliano} for reviews and see \cite{Krauss:1983zn,PhysRevLett.55.1422,Totani:1995rg,1996ApJ...460..303T,Malaney:1996ar,Hartmann:1997qe,PhysRevD.62.043001,Ando:2002zj,Fukugita:2002qw,Strigari:2003ig,Iocco:2004wd,Strigari:2005hu,Lunardini:2005jf,PhysRevD.72.103007,PhysRevC.74.015803,PhysRevD.79.083013,PhysRevLett.102.231101,PhysRevD.81.083001,PhysRevD.85.043011,Vissani:2011kx,Lunardini:2012ne,2013nakazatoimprint,Mathews:2014qba,Yuksel:2012zy,2015nakazatospectrum,Hidaka:2016zei,2017priya,2017horiuchi,2018moller,riya2020,2021kresse,ashida2022} for recent progress in DSNB predictions). The DSNB has not been detected yet, but recent upper flux limits at Super-Kamiokande (``Super-K'' or ``SK'') strongly disfavor optimistic models \cite{Super-Kamiokande:2002hei,Super-Kamiokande:2021jaq} and its confirmed detection is on the horizon \cite{Super-Kamiokande:2021the,2022yufengli}. Additional upcoming experiments like Hyper-Kamiokande \cite{Abe:2011ts}, JUNO \cite{JUNO:2015zny}, and DUNE \cite{DUNE:2015lol} will also probe the DSNB in the near future.

With only a single sample of supernova neutrino detections from 1987, DSNB predictions are informed primarily by the neutrino emission predicted by simulations (e.g., Refs.~\cite{2017priya,2018moller} and with extensive simulation sets, see Refs.~\cite{2017horiuchi,2021kresse}). Because of this, DSNB predictions are subject to inherent uncertainties of the simulations. 

In the last few years, there have been a great number of successful examples of fully three-dimensional (3D), robust CCSN simulations (see e.g., Refs.~\cite{2018summa,2019muller,2019burrows,2020stockinger,2021bollig,nakamura21,matsumoto21}). These capture the dynamic details that spherically symmetric 1D simulations inherently cannot. One example is turbulent processes like convection that affect the neutrino-driven explosion mechanism (see Ref.~\cite{2021burrows} and references therein). Another is the simultaneous mass accretion and explosion that can increase the neutrino luminosities, neutrino mean energies, and explosion energies compared to 1D simulations \cite{2006dessart,2017radice,2019burrows}. As a result, this ``accretion phase'' which occurs $\lesssim1\,{\rm s}$ is increasingly better understood by simulations in recent years.

Calculations post $\sim 1$ s, however, are less common and often limited by high computational costs. After $\sim1\,{\rm s}$, the cooling of the protoneutron star (PNS) becomes the dominant source of neutrinos. While progenitor dependent, typically $>50$\% of the total liberated neutrino energy is emitted in this phase. Moreover, these neutrinos are also important, e.g., for the determination of neutron star properties \cite{2019suwa,2021li,2022nakazato}. While studies of the cooling phase in the context of a nearby CCSNe exist (e.g., \cite{2019suwa,2021li,2022nakazato,2021li}), the extent to which they play a role in the DSNB has been explored much less. References~\cite{2017horiuchi,2021kresse} are two examples where extensive simulation sets inform the predicted DSNB neutrino emission, but rely on other methods for estimating the long-term cooling-phase neutrino emission. Reference~\cite{2017horiuchi} for example took simple analytic estimates for the PNS to estimate the energy liberated in neutrinos, while Ref.~\cite{2021kresse} used 1D simulations with calibrated engines.

\begin{table*}
\caption{\label{tab:overview}Overview of the simulation sets studied in this paper. Although this list is not exhaustive, it highlights several simulations done that include a cooling-phase component. Many are long enough so that the impact of this later phase on the DSNB can be studied. A check mark (\checkmark) means that we used that set in our analysis. ``D'' refers to the spatial dimension of the simulation. ``Duration(s)'' refers to the length postbounce of each simulation. Finally, ``Explosion model and cooling signal'' describes the method by which the explosion is induced and/or method for calculating the late neutrino signal. ``PNSC'' refers to protoneutron star cooling.}
\begin{ruledtabular}
\begin{tabular}{llllll}&
\textrm{Simulation set}&
\textrm{D}&
\textrm{Duration(s)}&
\textrm{EOS}&
\textrm{Explosion model and cooling signal}\\
\colrule
     &Fischer \textit{et al}. (2009) \cite{2010fischer}&1D&$\approx10$&Shen&Enhance electronic charged current rates\\
     (\checkmark)&Nakazato \textit{et al}. (2013) \cite{2013nakazato}&1D&20&Shen&Assumed transition time to PNSC simulation\\
     (\checkmark)&H{\"u}depohl (2014) \cite{2014PhDT.......436H}&1D&$\approx10$&Shen/LS220&Artificially decrease density\\
     &Sukhbold \textit{et al}. (2016) \cite{2016sukhbold}&1D&$\approx1$&LS220&Calibrated central engine \cite{2012ugliano}\\
     &Summa \textit{et al}. (2016) \cite{2016summa}&2D&$\approx1$&LS220&Self-consistent shock revival\\
     (\checkmark)&Horiuchi \textit{et al}. (2018) \cite{2017horiuchi}&2D&100&LS220&Self-consistent shock revival, late-time analytic\\
     (\checkmark)&Burrows \textit{et al}. (2019) \cite{2019burrows}&3D&$\approx1$&SFHo&Self-consistent shock revival\\
     (\checkmark)&Sumiyoshi \textit{et al}. (2019) \cite{2019sumiyoshi}&1D&$\approx60$&TM1/TM1e&Hydrodynamic simulations set initial conditions for PNSC\\
     (\checkmark)&Suwa \textit{et al}. (2019) \cite{2019suwa}&1D&$>50$&TM1&Consistent cooling, connect to Nakazato \textit{et al}. (2013)\\
     &Li \textit{et al}. (2021) \cite{2021li}&1D&$\approx100$&Schneider&Replace outer layers with pressure boundary\\
     (\checkmark)&Bollig \textit{et al}. (2021) \cite{2021bollig}&3D&$\approx7$&LS220&Self-consistent shock revival, connect to 1D sim at $\sim1.7\,{\rm s}$\\
     &Nagakura \textit{et al}. (2021) \cite{2021nagakura}&2D&$\approx4$&SFHo&Self-consistent shock revival\\
\end{tabular}
\end{ruledtabular}
\end{table*}

The primary focus of this study is to implement and compare different methods for estimating the cooling-phase neutrino emission, in order to quantify how the cooling phase impacts the DSNB signal. We find that, within reasonable variations of cooling-phase estimates, the predicted DSNB rate can vary by a factor of $\sim 2$--3. Recently, SK has been enhanced with gadolinium salt (``SK-Gd'') which will allow for neutron tagging of true DSNB events and reject backgrounds \cite{2004beacom,Super-Kamiokande:2021jaq}. This dramatically improves the detection prospects of the DSNB in the next decade. Our study suggests that improving the understanding of the late-phase neutrino emission will be important for the uncertainties in the DSNB as we enter the SK-Gd era.

This paper is organized as follows. In Sec.~\ref{simulationset} we describe the simulation data we use in our study. In Sec.~\ref{latephasestrategies} we describe the different methods for estimating the cooling phase of neutrino emission. In Sec.~\ref{latecomparison} we check the validity of these strategies against the results of a 3D simulation. In Sec.~\ref{sec:dsnbevents} we give quantitative DSNB event rates at SK-Gd. Finally, in Sec.~\ref{sec:discussion} we discuss our framework and summarize how our late-phase strategies lead to a large difference in DSNB rates.

\section{\label{sec:latephase}Characterizing the Cooling Phase}

\subsection{Simulation survey}\label{simulationset}

We first summarize the core-collapse simulation sets used in our study, including studies of the cooling phase, but also the accretion phase and collapse to black holes.

In Table~\ref{tab:overview} we highlight a selection of recent studies of core collapse into the PNS cooling phase using a variety of techniques spanning spatial dimensionality, nuclear equation of state (EOS), and implementation for shock revival. Here, the EOS includes Shen \cite{1998shen}, LS220 \cite{1991NuPhA.535..331L}, SFHo \cite{2013steiner}, TM1/TM1e \cite{2011shen}/\cite{2020ApJ...891..148S} (where TM1 is an updated version of the Shen EOS), and Schneider \cite{2020schneider}. These simulations do not make up an exhaustive list but begin to reveal how spatial dimension, EOS, implementation details, and perhaps artificial biases lead to systematic differences.

\begin{figure}
\includegraphics[width=\linewidth]{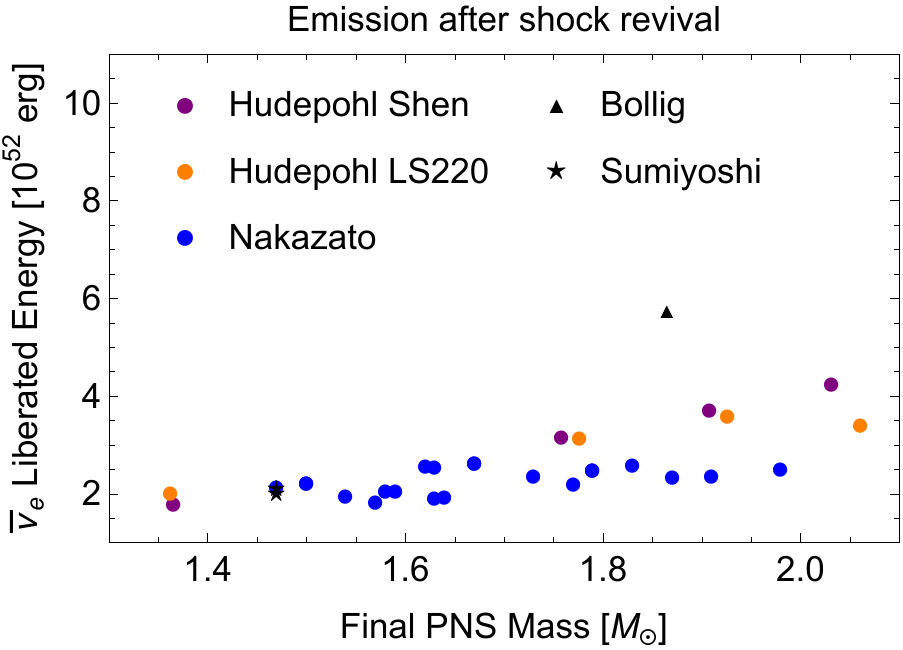}
\includegraphics[width=\linewidth]{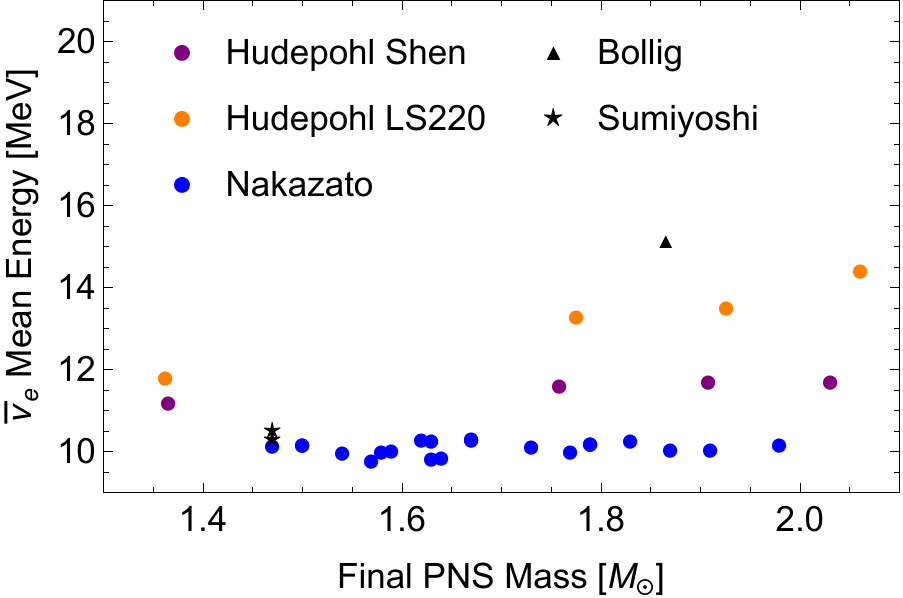}
\caption{Comparison of the time-integrated $\overline{\nu}_e$ spectral parameters in the cooling phase of PNS evolution shown against the final baryon mass of the PNS. Here we compare those quantities of the H{\"u}depohl \cite{2014PhDT.......436H}, Nakazato \cite{2013nakazato}, Bollig \cite{2021bollig}, and Sumiyoshi \cite{2019sumiyoshi} (integrated from shock revival time to the end of simulation time: $\sim10$, $20$, $\sim7$, and $\sim60\,{\rm s}$, respectively). Their EOSs are the Shen/LS220, Shen, SFHo, and TM1/TM1e, respectively. See text and Table~\ref{tab:overview} for details. There are other differences not mentioned here, but these serve to show some dependence of neutrino emission on final PNS mass and hint at systematic differences due to e.g. EOS. More simulation details can be found in Table~\ref{tab:overview}}
\label{fig:integratedlates}
\end{figure}

We show the time-integrated $\overline{\nu}_e$ liberated energy (top panel) and the $\overline{\nu}_e$ mean energies (bottom panel) in Fig.~\ref{fig:integratedlates}, for a subset of core-collapse simulations in Table~\ref{tab:overview}. We compare the neutrino emission properties accounting for the final baryon mass of the PNS. In general, we confirm previously found trends of increased neutrino liberated energy and mean energy as the final PNS mass is increased (e.g., \cite{2013nakazato}). However, different strategies introduce systematic differences. For example, the EOS plays a big role in the mean energy. We can see this from comparing the Shen and LS220 simulations by the ``H{\"u}depohl'' study \cite{2014PhDT.......436H}. Also, the ``Bollig'' \cite{2021bollig} simulation, which is spatially three dimensional until $\sim1.7\,{\rm s}$ before connecting to a 1D simulation, point to higher neutrino energetics and mean energies than H{\"u}depohl \cite{2014PhDT.......436H}, ``Nakazato'' \cite{2013nakazato}, and ``Sumiyoshi'' \cite{2019sumiyoshi} which are fully simulated in spherical symmetry. While this seems to be driven in part by aspherical mass infall and ouflows, conclusive statements cannot be made with only one simulation comparison. Finally, we can see that the Sumiyoshi simulations using the TM1 and TM1e EOS result in very similar integrated neutrino emission to those of Nakazato, and both yield significantly lower liberated and mean energies when compared to other 1D simulations. This could arise from differences in how the cooling phase is initiated and/or in more subtle implementation details like numerical methods, resolution, and neutrino interactions.

For the DSNB we also need neutrino emission from the early hydrodynamic phase of core-collapse evolution which precedes the cooling phase. For this, we use the angle-averaged three-dimensional simulation data from Ref.~\citep{2019burrows} (referred to as ``Burrows'' hereafter, see also Ref.~\cite{Nagakura:2020qhb} for discussion of the neutrino emission). This simulation set includes more than a dozen progenitors each simulated until close to $\sim 1\,{\rm s}$ postbounce. We exclude, however, the $13,~14$ and $15\,M_{\odot}$ progenitors since they do not explode by the end of simulation run-time. We augment the Burrows set with the neutrino signal from Ref.~\cite{2010PhRvL.104y1101H}, an electron-capture supernova of an ONeMg core.

Finally, we also want to account for the neutrino emissions from core collapses directly to black holes, i.e., failed CCSNe. To do this, we take a similar method to that of Ref.~\cite{2018moller} in their conservative scenario. More specifically, they assume all progenitors with initial masses $>40\,M_{\odot}$ become black holes ($\sim$8.4\% of core collapse) and adopt the neutrino signal based on two simulations, the ``s40'' and ``s40s7b2'' models of Ref.~\cite{2014PhDT.......436H}, both using the LS220 EOS. There are also cases of fallback black hole formation where material falls back onto the PNS after shock revival \cite{2021li}, but we ignore their contribution since they are estimated to be rare \cite{2016ertl,2016sukhbold,Janka_2017}

\subsection{Late-phase strategies}\label{latephasestrategies}

Here we estimate the neutrino emission from the late cooling phase of PNS evolution with several strategies. We discuss five estimates using four strategies. In the next section, we will explore how they impact the DSNB.

\subsubsection{Constant mean energy method}\label{2018method}

The simplest of our strategies is a simple analytic treatment. First, as was done in Ref.~\cite{2017horiuchi}, we assume that the mean energy is constant after the hydrodynamic simulation concludes. Next, to estimate the energy liberated, we again follow Ref.~\cite{2017horiuchi} and assume that all of the remaining gravitational binding energy released after simulation is released as neutrinos. This requires the evolution of the PNS mass, which is taken from Ref.~\cite{Arcones:2006uq} as $M(t)=M_0+M_1(1-e^{-t/t_M})$, where the parameters $M_0$, $M_1$, and $t_M$ are found by fitting to the PNS mass evolution in the simulation. However, we adopt a final PNS radius of 12 km, rather than 15 km in Ref.~\cite{2017horiuchi}, which is more consistent with the SFHo EOS. With both of these final parameters, we can calculate the binding energy after the simulation.

In reality, the mean energy decreases in the cooling phase. Thus, the results we calculate from the constant mean energy method should be seen as upper limits.

\subsubsection{Analytic solution method}\label{suwaform}

In Ref.~\cite{2020suwa}, analytic solutions of the neutrino luminosity and mean energy of PNS cooling are derived assuming spherical symmetry and a thermal energy spectrum. We take the one-component functional form from Ref.~\cite{2020suwa} to estimate the luminosity and mean energy after $\sim 1\,{\rm s}$. As input parameters, these functions require the final PNS baryonic mass, radius, and the total liberated energy. The analytic method is also dependent on two additional parameters: $g$ and $\beta$, the density correction and opacity boosting factors which are used in Ref.~\cite{2020suwa} as effective parameters to parametrize the PNS differences from the Lane-Emden structure and model the increased scattering due to heavy nuclei, respectively. 

\begin{figure}
\includegraphics[width=\linewidth]{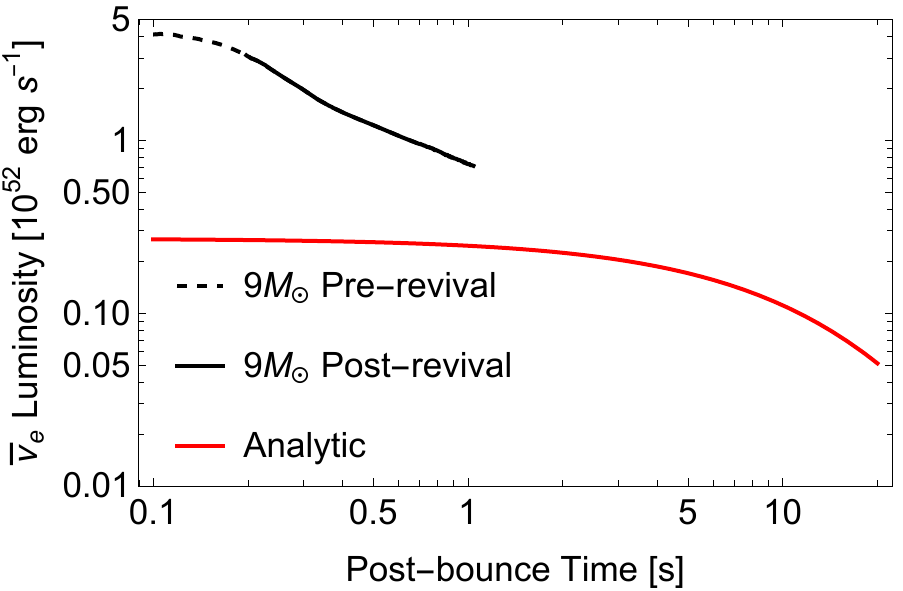}
\includegraphics[width=\linewidth]{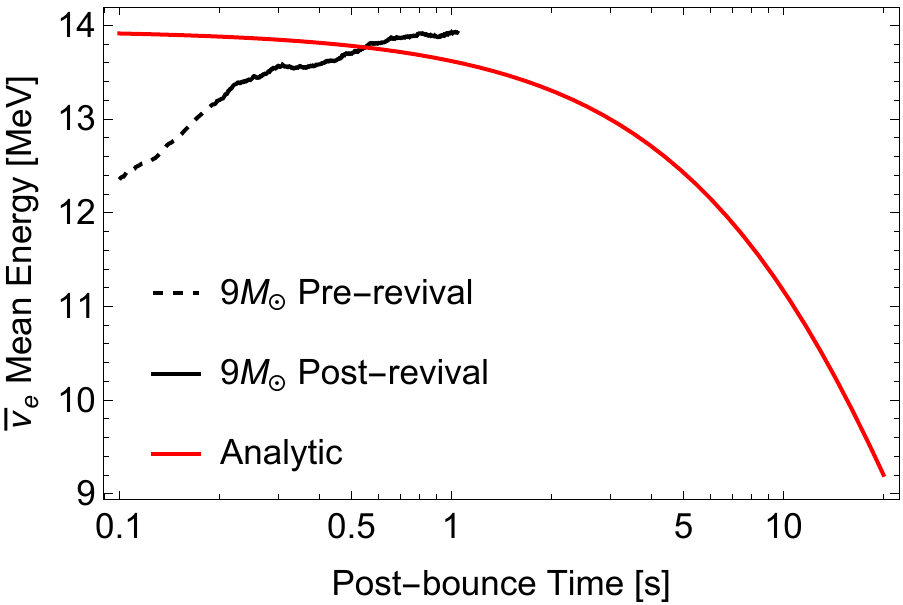}
\caption{The functional form from Ref.~\cite{2020suwa} (red) using the effective density correction $g$, and opacity boosting factor $\beta$, compared to the hydrodynamic neutrino data for the $9\,M_{\odot}$ progenitor (black). These parameters are chosen to best agree with the mean energy data and then are used in the luminosity function. Around the end of the simulation time, the functional form underpredicts the neutrino luminosity and mean energy. Later comparisons show that integrating these functions still gives reasonable results, but ideally would be applied to longer-duration ($\gtrsim10\,{\rm s}$) data.}
\label{fig:suwasols}
\end{figure}

We aim to add the analytic cooling solutions to the ends of the Burrows simulations. We take the final mass as $\min[M_{\rm end},M_{\rm max}]$, where $M_{\rm end}$ is the PNS mass at the end of the simulation and $M_{\rm max}$ is the maximum NS mass from the SFHo EOS \cite{2013steiner}. We then use the SFHo mass-radius relationship to estimate the final radius, giving us enough information to estimate the remaining gravitational binding energy. As was done in Ref.~\cite{2020suwa}, to estimate the effective parameters we can either fit to the neutrino luminosity or neutrino mean energy or both. We opt to fit to the mean energy, since it varies less over this accretion phase and the luminosity curve is always below the simulation data. We use a sum-of-least-squares method on the simulation mean energy data after each progenitor's revival time and find best-fit values. For all progenitors in Burrows we set a default value of $g=0.07$ and allow $\beta$ to vary, since the analytic functions depend only on the product $g\times\beta$. This returns values of $\beta$ between $\sim20$--40 which are within the target range indicated by \cite{2020suwa}. 

Figure~\ref{fig:suwasols} shows the result of this best-fitting procedure for the $9\,M_{\odot}$ progenitor. The black line shows the Burrows simulation data, which we split into prerevival (dashed) and postrevival (solid), the latter of which is used to determine the analytic (red) effective parameter. Note that we show the analytic model down to $\sim 0.1\,{\rm s}$ but only use it beyond the available simulation data. Note also that a consequence of fitting to the mean energy is the poorer fit to the neutrino luminosity. However, integrating this luminosity up to long times (e.g., $20\,{\rm s}$) produces reasonable results, i.e., agrees well with the gravitational binding energy liberated (see next sections). 

Although other progenitors in Burrows show qualitatively similar results, the $9\,M_{\odot}$ progenitor shows an earlier shock revival among the Burrows set and hence the analytic mean energy solution resembles the data more closely around the end of the simulation time. However, the neutrino mean energy is still increasing in the simulation after the shock revives and it is not certain when the transition to the true cooling phase of the PNS occurs. Depending on the true evolution of the mean energy beyond the available simulation, this method, then, could still under- or overestimate the true mean energy. Given this uncertainty, we caution that this method would be more effective when simulation data are available out to $\gtrsim10\,{\rm s}$ or until accretion luminosity has sufficiently reduced.

\subsubsection{Correlation method}\label{nakazatocorrelation}

\begin{figure}
\includegraphics[width=\linewidth]{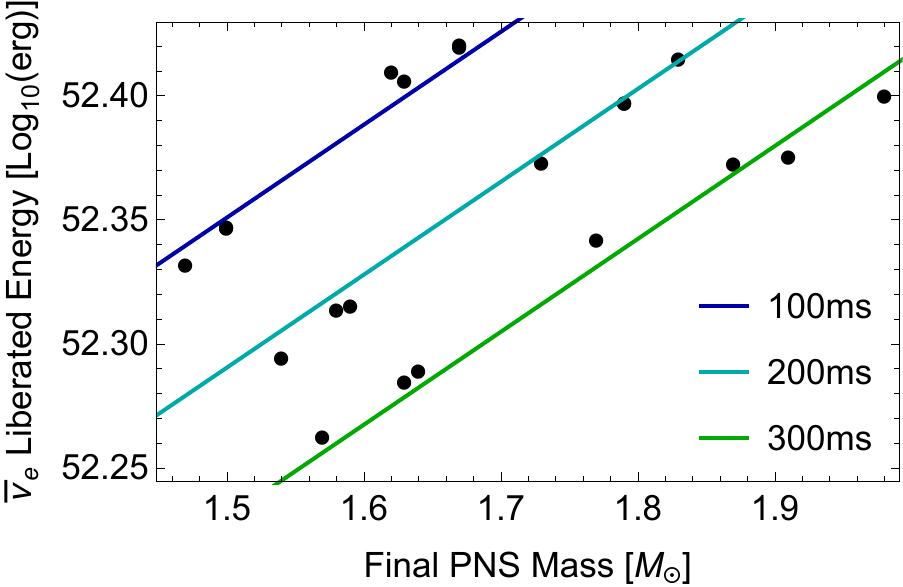}
\includegraphics[width=\linewidth]{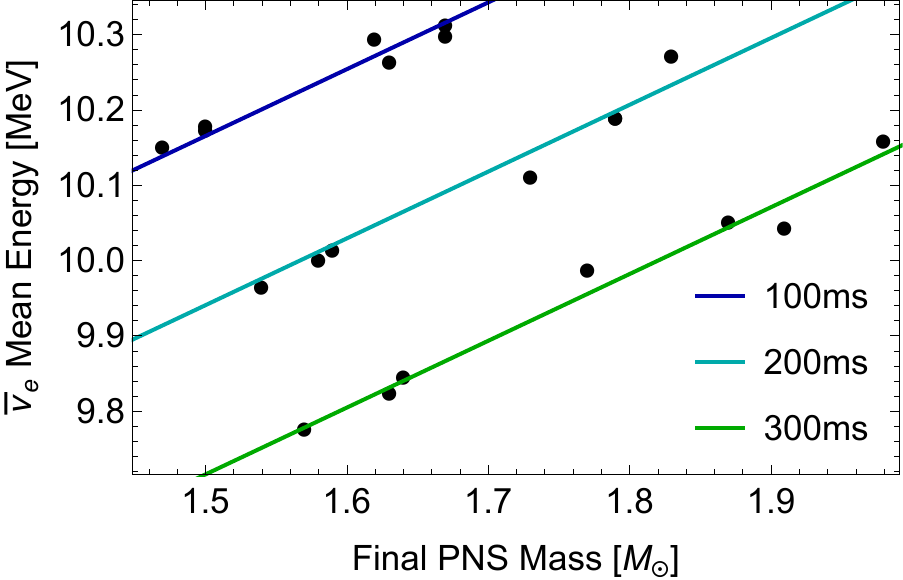}
\caption{Time-integrated (from shock revival time to $20\,{\rm s}$) neutrino spectral parameters (top panel, $\bar{\nu}_e$ energy liberated; bottom panel, $\bar{\nu}_e$ mean energy) from the Supernova Neutrino Database \cite{2013nakazato}. Approximately linear trends are visible with respect to the PNS final baryon mass (x-axis) and shock revival time (100, 200, and 300 ms as labeled); see Eqs. (\ref{etotcorrfit}) and (\ref{emeancorrfit}). With these linear trends, we can estimate the late-phase neutrino emission for combinations of shock revival and final PNS mass.}
\label{fig:ncorrelation}
\end{figure}

For our next strategy, we use the Supernova Neutrino Database of Ref.~\cite{2013nakazato} and examine properties of the late-phase spectral parameters. This database is particularly valuable as a reference of long-term simulations that are carried out for $20\,{\rm s}$ post-core-collapse for 21 progenitors. In Refs.~\cite{2013nakazatoimprint,2015nakazatospectrum}, a correlation was found between neutrino emission, shock revival time, and PNS mass. Further, it was shown that this dependence can leave an imprint on the DSNB signal. We confirm this correlation, and also quantify a new linear relationship of the logarithm of energy liberated and mean energy with the shock revival time and final PNS baryon mass for the cooling phase. We show these relationships in Fig.~\ref{fig:ncorrelation} for $\overline{\nu}_e$ neutrinos for all progenitors of the Supernova Neutrino Database.

It should be noted that the simulations of Ref.~\cite{2013nakazato} are 1D and do not attempt to tune the models to explode. Rather, the revival time is \textit{ad hoc}, set by an assumed transition time from accretion hydrodynamical simulation to the PNS cooling simulation. The simulations of Ref.~\cite{2013nakazato} also use the Shen EOS \cite{1998shen}. This limits the range of final masses and, thus, the range of applicability of this method. It should, in principle, turn out that other EOSs lead to modifications to this method, including the range of final masses that it applies to. We discuss this and account for some of this in the next section, Sec.\ref{hurenorms}.

\begin{table}[b]
\caption{\label{tab:ncorrelations}Table of Nakazato correlations [Eqs. (\ref{etotcorrfit}) and (\ref{emeancorrfit})]: This is the ``Corr'' method, where $M_{\rm fin}$ is the final baryonic mass, $t_{\rm rev}$ is the shock revival time, and $\alpha,~\beta,~\textrm{and}~\gamma$ are the coefficients. }
\begin{ruledtabular}
\begin{tabular}{ccccc}
&Flavor $i$&
\textrm{$\alpha$}&
\textrm{$\beta$}&
\textrm{$\gamma$}\\
\colrule
&$\nu_e$&$0.317$&$-5.54\times 10^{-4}$&$51.92$\\
$\textrm{Log}_{10}(E)$&$\overline{\nu}_e$&$0.375$&$-6.04\times 10^{-4}$&$51.85$\\
&$\nu_x$&$0.412$&$-3.97\times 10^{-4}$&$51.87$\\
\hline
&$\nu_e$&$0.999$&$-1.17\times 10^{-3}$&$6.99$\\
$\epsilon$&$\overline{\nu}_e$&$0.887$&$-2.25\times 10^{-3}$&$9.06$\\
&$\nu_x$&$0.892$&$-2.36\times 10^{-3}$&$10.1$\\
\end{tabular}
\end{ruledtabular}
\end{table}

In Table \ref{tab:ncorrelations} we show the linear fits to these data for all neutrino flavors for the logarithm of liberated energy first, then the mean energies. These fits are of the form
\begin{eqnarray}
{\rm Log}_{10}(E_i)&=&\alpha_i^{(E)} M_{\rm fin} + \beta_i^{(E)} t_{\rm rev} + \gamma_i^{(E)},
\label{etotcorrfit} \\
\epsilon_i&=&\alpha_i^{(\epsilon)} M_{\rm fin} + \beta_i^{(\epsilon)} t_{\rm rev} + \gamma_i^{(\epsilon)}, \label{emeancorrfit}   
\end{eqnarray}
where $M_{\rm fin}$ is the final baryonic mass of the PNS, $t_{\rm rev}$ is the shock revival time, and $\alpha,~\beta$ and $\gamma$ are the fit coefficients for neutrino flavor $i$. Since the cooling data from the Supernova Neutrino Database are computed from shock revival time to $20\,{\rm s}$, this method yields the time-integrated neutrino emission for the cooling phase until $\sim 20$ s. We only show results in this section for $\overline{\nu}_e$, but other flavors show similar trends; see the Appendix, Figs.~\ref{fig:ncorrelationfore} and \ref{fig:ncorrelationforx}.

\subsubsection{Renormalized correlation method}\label{hurenorms}

\begin{figure}
\includegraphics[width=\linewidth]{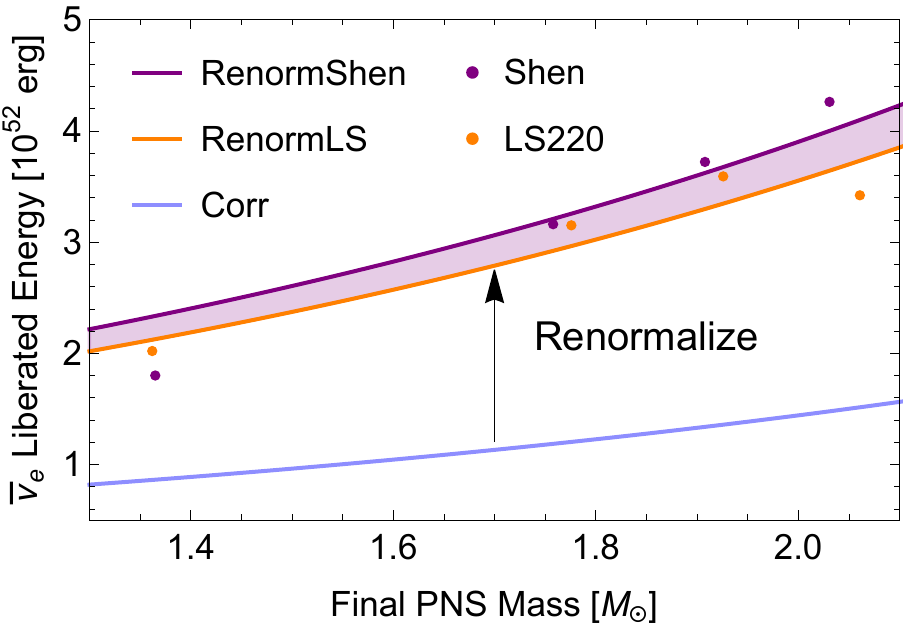}
\includegraphics[width=\linewidth]{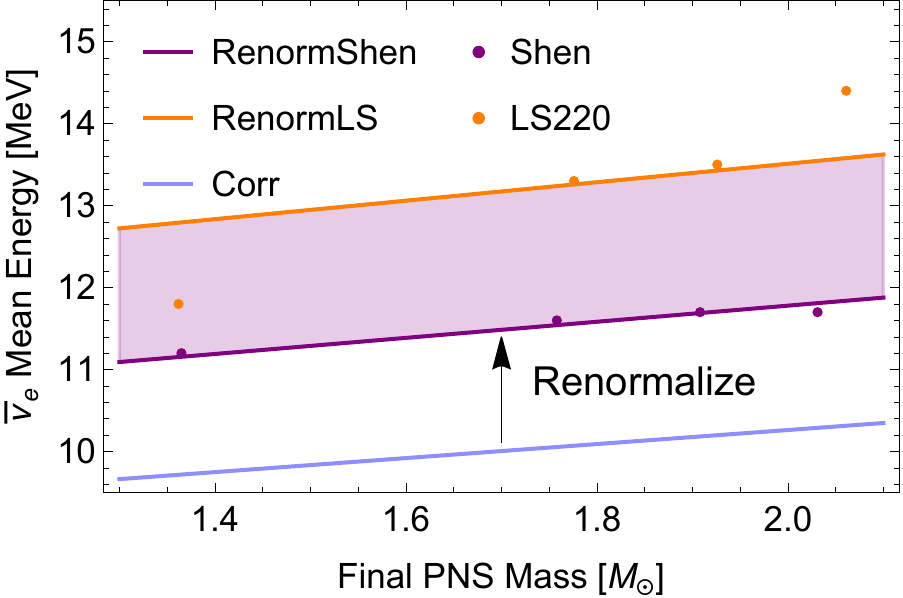}
\caption{To properly compare these neutrino spectral parameters, we time integrate the liberated and mean energies from shock revival time to $\sim10\,{\rm s}$. We then renormalize the $\sim10\,{\rm s}$ revival-time--final-mass correlations using a sum-of-least-squares method to best fit the four H{\"u}depohl progenitors. This is performed for the Shen and LS220 EOSs to capture some EOS and simulation dependence on the integrated neutrino spectral parameters.}
\label{fig:hurenorms}
\end{figure}

One-dimensional CCSN simulations, with some exceptions, require artificial induction of explosion, and the strategy for inducing explosions introduces model dependence. Furthermore, the EOS will impact the evolution of the PNS cooling phase. To cover these and other range of physics possibilities, we make use of additional cooling-phase simulations by Ref.~\cite{2014PhDT.......436H}. In these H{\"u}depohl simulations, a small suite of spherically symmetric simulations are carried out to $\sim10\,{\rm s}$ for both Shen and LS220 EOS. For our purposes, we use their standard simulations which do not use a mixing-length scheme to model multidimensional dynamics and corrections to neutrino opacities. However, Ref.~\cite{2014PhDT.......436H} considered only four progenitors, less than the sample studied by Ref.~\cite{2013nakazato} discussed in the previous section and not enough to robustly extract a trend. Therefore, we compare outcomes of Refs.~\cite{2013nakazato,2014PhDT.......436H}, and explore how the simulation dependence and progenitor dependence can be incorporated by a renormalization factor.

To this end, we first integrate the simulations of \cite{2013nakazato} out to $10\,{\rm s}$, i.e., comparable to the duration of the simulations of Ref.~\cite{2014PhDT.......436H}, for a fair comparison. We then take the overall normalization as a free parameter and perform a sum-of-least-squares fitting procedure to the four simulations of Ref.~\cite{2014PhDT.......436H}, assuming a revival time of $500\,{\rm ms}$ (this is when the explosion is artificially induced in the H{\"u}depohl simulations). In other words, we calculate $E_i'=\mathcal{N}_iE_i$ and $\epsilon_i'=\mathcal{N}_i\epsilon_i$ where $E_i$ and $\epsilon_i$ are defined in Eqs. (\ref{etotcorrfit}) and (\ref{emeancorrfit}), $\mathcal{N}_i$ is the overall normalization parameter for neutrino flavor $i$, and this is done for both the Shen and LS220 EOSs. Note that we found the final mass-revival time correlation with the logarithm of energy liberated, but we renormalize the energy liberated linearly.

The results of this procedure are shown in Fig.~\ref{fig:hurenorms}, where the blue curves are the original trends of \cite{2013nakazato} and the orange and purple curves are the renormalized curves to the LS220 and Shen EOS simulations of \cite{2014PhDT.......436H}, respectively. Interestingly, the original trend of \cite{2013nakazato} shows a remarkably good description of the \cite{2014PhDT.......436H} simulations, as seen by how well the renormalized curves fit through the simulation points. The comparison also highlights the large impact the EOS plays on the neutrino average energy (and much less for the liberated energy). This large mean energy difference ends up playing an important role in the predicted DSNB events, as we see in later sections. We only show results in this section for $\overline{\nu}_e$, but other flavors show similar trends; see the Appendix, Figs.~\ref{fig:hurenormsfore} and \ref{fig:hurenormsforx}. Finally, in Table~\ref{tab:renorms}, we show the renormalization constants, $\mathcal{N}_i$, to the Corr strategy.  

As with the previous correlations, we construct the renormalized form for the integrated neutrino spectral parameters from shock revival time to $20\,{\rm s}$. This constitutes our final late-time strategy.

\begin{table}[b]
\caption{\label{tab:renorms}Table of H{\"u}depohl renormalizations for liberated and mean neutrino energies. In the RenormShen and RenormLS methods, we renormalize the liberated and mean energy correlations using the same $\alpha$, $\beta$, and $\gamma$ coefficients from the Corr method (see Table~\ref{tab:ncorrelations}).}
\begin{ruledtabular}
\begin{tabular}{cccc}
&Flavor $i$&Shen $\mathcal{N}$&LS220 $\mathcal{N}$\\
\colrule
&${\rm \nu_e}$&2.57&2.33\\
$E'$&${\rm \overline{\nu}_e}$&2.71&2.46\\
&${\rm \nu_x}$&1.72&1.63\\
\hline
&${\rm \nu_e}$&1.13&1.28\\
$\epsilon'$&${\rm \overline{\nu}_e}$&1.15&1.32\\
&${\rm \nu_x}$&1.00&1.16\\
\end{tabular}
\end{ruledtabular}
\end{table}

\subsection{Comparison to 3D simulation}\label{latecomparison}

To make sure our late-phase strategies return reasonable results, we test them against the Bollig simulation~\cite{2021bollig}, which extends a 3D hydrodynamic simulation with a 1D cooling simulation out to $\sim 7\,{\rm s}$ postbounce. We first estimate the time-integrated luminosity and mean energy up to the end of the simulation. We then compare these with the values calculated from the five estimates. The results are shown in Table~\ref{tab:bolligcomparison}. 

Since all of the strategies are intended to estimate the neutrino spectral parameters after $\sim 20\,{\rm s}$ postbounce, we have to slightly modify the strategies to instead estimate these parameters until $\sim 7\,{\rm s}$ postbounce. For ``Const,'' we do not modify the estimation for liberated energy since the PNS mass and radius do not change much between $7$ and $20\,{\rm s}$. However, to give a more reasonable comparison, we take the mean energy to be constant after $\sim1.7\,{\rm s}$ (the beginning of the 1D neutrino signal for the Bollig simulation) instead of the end of the simulation. For ``Analyt,'' we modify the method in the following way: We find the best-fit $g$ and $\beta$ parameters by finding the minimum sum of least squares in the time range between $\sim1.7$ and $\sim7\,{\rm s}$. To calculate the neutrino spectral parameters, we then integrate the Bollig and analytic solutions up to $7\,{\rm s}$. Finally, for Corr and ``RenormShen/LS'' we find and apply correlations integrated to $7\,{\rm s}$ instead of $20\,{\rm s}$, and adopt the time when the shock radius reaches 400 km as the shock revival time. 

These Bollig simulation data are an interesting test case for the Analyt method because of how long this simulation is carried out. We found that low $g\times\beta$ values fit the mean energy curve best by eye, but these low values may not be physically appropriate for the late-time solutions \cite{2020suwa}. The sum-of-least-squares method of finding $g\times\beta$ values return integrated neutrino spectral parameters that agree fairly well with Bollig spectral parameters, but do not resemble the mean energy and luminosity curves well. This likely is a result of the continued mass accretion postshock revival. Reference~\cite{2020suwa} also describes a two-component approach. This does a slightly better job than the one-component solution, but we use the results of the one-component solution in Table~\ref{tab:bolligcomparison} since this is the strategy we take when applying this method to the Burrows simulation data.

Overall, we find the strategies provide reasonable estimates for the Bollig liberated and mean energies, perhaps with the exception of the Corr where the liberated energy is notably lower. In liberated energy, the Const method is slightly higher than the simulation data since this is closer to estimates of the total gravitational binding energy released over the entire PNS evolution. The Analyt method and the renormalized correlations slightly underpredict the simulation data. In mean energy, the Const method understandably overpredicts the mean energy since the value does not reflect the PNS cooling that occurs at later times. The other strategies slightly underpredict the neutrino mean energy at the end of simulation, but are especially close for the Analyt and RenormLS strategies.

\begin{table}[b]
\caption{\label{tab:bolligcomparison}
A comparison of the integrated spectral parameters: liberated and mean energies for antielectron neutrinos. Bollig values represent values until the end of their simulation $\sim7\,{\rm s}$ \cite{2021bollig} and the strategies have been modified to estimate these parameters up to the same time, instead of $20\,{\rm s}$ as intended.}
\begin{ruledtabular}
\begin{tabular}{ccc}
Strategy&$E_{\rm{\overline{\nu}_e}}~(10^{52}\,{\rm erg})$&$\epsilon_{\overline{\nu}_e}~({\rm MeV})$\\
\colrule
\textrm{Bollig numerical}&7.65&14.82 \\
\hline
\textrm{Const}&8.93&15.19\\
\textrm{Analyt}&5.54&14.74\\
\textrm{Corr}&3.14&12.38\\
\textrm{RenormShen}&6.90&12.61\\
\textrm{RenormLS}&6.44&13.96\\
\end{tabular}
\end{ruledtabular}
\end{table}

\subsection{Application to simulation suite}\label{lateapplication}

\begin{figure}
\includegraphics[width=\linewidth]{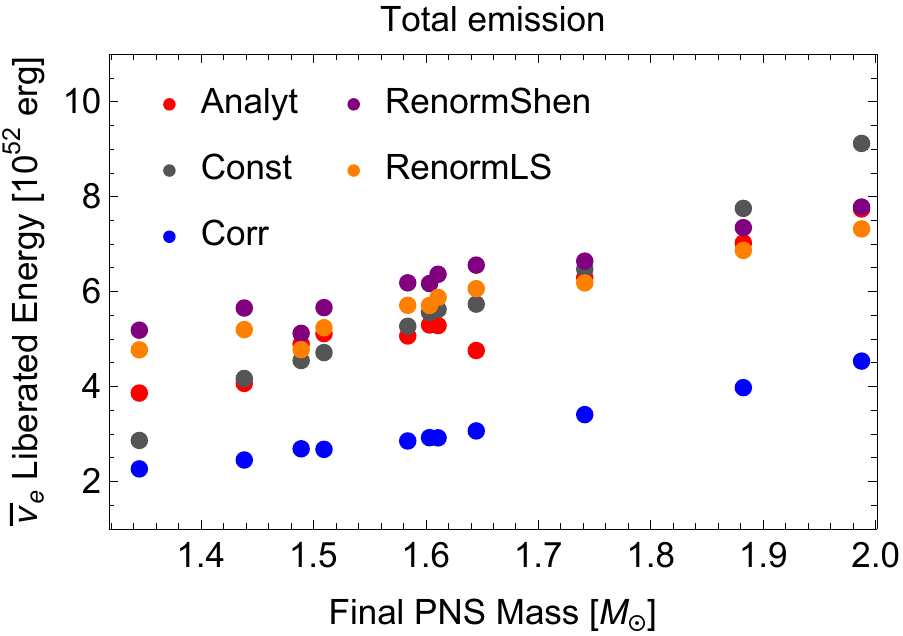}
\includegraphics[width=\linewidth]{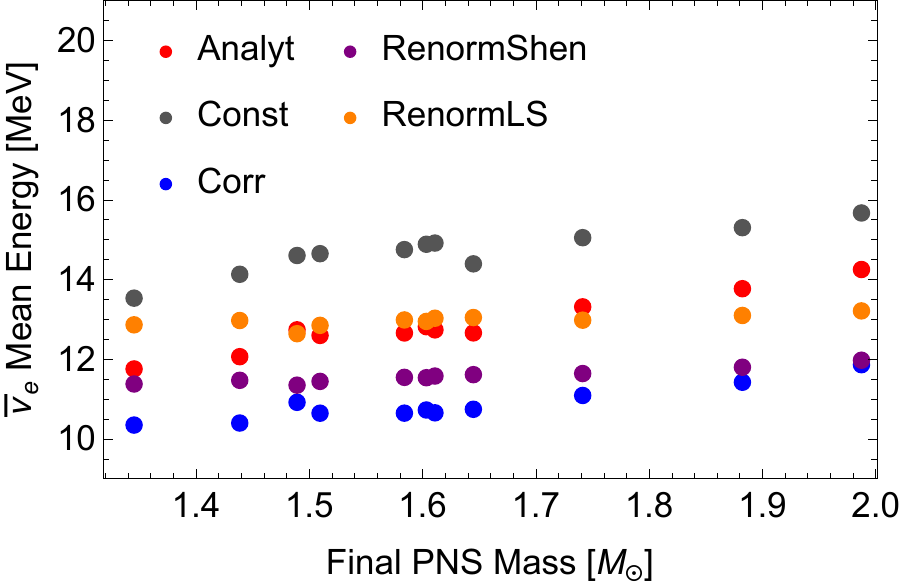}
\caption{Time-integrated (represents integration from $0$ to $\sim20\,{\rm s}$ postbounce) neutrino spectral parameters using the Burrows simulation set for the hydrodynamic phase and five estimates, as labeled, for adding the cooling phase. In energy liberated, Corr is systematically lower than the other four estimates. In mean energy, there is a fairly clear hierarchy where Const is always highest and Corr lowest. For each method, the early-phase neutrino emission is calculated using the SFHo EOS, but in the Corr and RenormShen methods the late-phase neutrino emission is dependent on the Shen EOS and the RenormLS method is dependent on the LS220 EOS.}
\label{fig:fivelate}
\end{figure}

In this section we apply our strategies to the hydrodynamic simulation data of the Burrows set (see Sec.~\ref{simulationset}) and compare the outcomes. We show in the top panel of Fig.~\ref{fig:fivelate} the liberated energies. These are all quite similar and do not show any systematic preferences by strategy, with the exception of Corr which is systematically lower than the others by a factor $\sim 2$; this is consistent with the check against the Bollig simulation (see previous section). Interestingly, the comparison shows how different EOSs (Shen vs LS220) do not yield large differences in the total liberated energy.

Mean energies, however, show a clear spread in strategies. Unsurprisingly, Const returns the highest mean energies; in this method, the mean energy is kept fixed to the end of the hydrodynamical simulation and neglects the reduction during cooling. On the other end, Corr gives the lowest mean energies. Between these are the results renormalized by the simulation set of \cite{2014PhDT.......436H} (where RenormShen uses the same Shen EOS as Corr) and the results of the analytic solution method. As we will show in Sec.~\ref{numbers}, the mean energy still leads to a large impact on the predicted DSNB, and highlights the importance of quantifying the neutrino mean energy of the late phase.

\section{DSNB Event Numbers}\label{sec:dsnbevents}

\subsection{Predicting the DSNB}\label{prediction}

In order to predict the DSNB rate, we need the mean neutrino emission spectrum and the occurrence rate of core collapses. Using the integrated neutrino spectral parameters (the liberated and mean energies from our cooling phase strategies), we estimate the neutrino energy distribution with a pinched Fermi-Dirac distribution \cite{2003keil}:
\begin{equation}
    f(E)=\frac{(1+\alpha)^{1+\alpha}}{\Gamma(1+\alpha)}\frac{E_\nu E^{\alpha}}{(\epsilon_\nu)^{2+\alpha}}\exp\left[-(1+\alpha)\frac{E}{\epsilon_\nu}\right],
\end{equation}
where $\alpha$ is a shape parameter (sometimes called the pinching parameter), $E_{\nu}$ is the total liberated energy for neutrino $\nu$, and $\epsilon_{\nu}$ is the mean energy for neutrino $\nu$. We will consider separate neutrino emission spectra from successful and failed CCSNe since both are important for the DSNB.

To estimate the mean neutrino emission from a population of stars, we perform a weighted mean of stars based on the initial mass function (IMF). The IMF-weighted average neutrino spectrum is given:
\begin{equation}
    \frac{dN}{dE}=\sum_i\frac{\int_{\Delta M_i}\psi(M)dM}{\int_{M_0}^{M_f}\psi(M)dM}f_i(E)
\end{equation}
where $\Delta M_i$ is the mass range of mass bin $i$ and $\psi(M)=dn/dM$ is the IMF. We use the IMF $\psi(M)\propto M^{\eta}$ where $\eta=-2.15$ from Ref.~\cite{2003bg}. Here we take $M_0=8 M_\odot$ and $M_f=100 M_\odot$. On the low mass end of the IMF, core collapses of ONeMg cores (or ``electron-capture SNe'') make up a significant fraction of CCSNe, so we include a contribution from the $8.8\,M_\odot$ progenitor of Ref.~\cite{2010PhRvL.104y1101H}. In the range of intermediate masses, we take the progenitors used by Burrows. Finally, following Ref.~\cite{2018moller}, we conservatively represent the black hole (BH) channel by assuming progenitors with initial masses above $40\,M_\odot$ fail as CCSNe. For the ONeMg progenitor, we take $\Delta M_i=[8\,M_{\odot},~8.9\,M_{\odot}]$, for intermediate bins $\Delta M_i=[(M_{i-1}+M_i)/2,~(M_i+M_{i+1})/2]$, for our $25\,M_{\odot}$ progenitor, we take $\Delta M_i=[22.5\,M_{\odot},~40\,M_{\odot}]$, and for our BH channel we take mass bin $\Delta M_i=[40\,M_{\odot},~100\,M_{\odot}]$, where $M_i$ is the initial mass. In total, we have 12 mass bins.

\begin{figure}
\includegraphics[width=\linewidth]{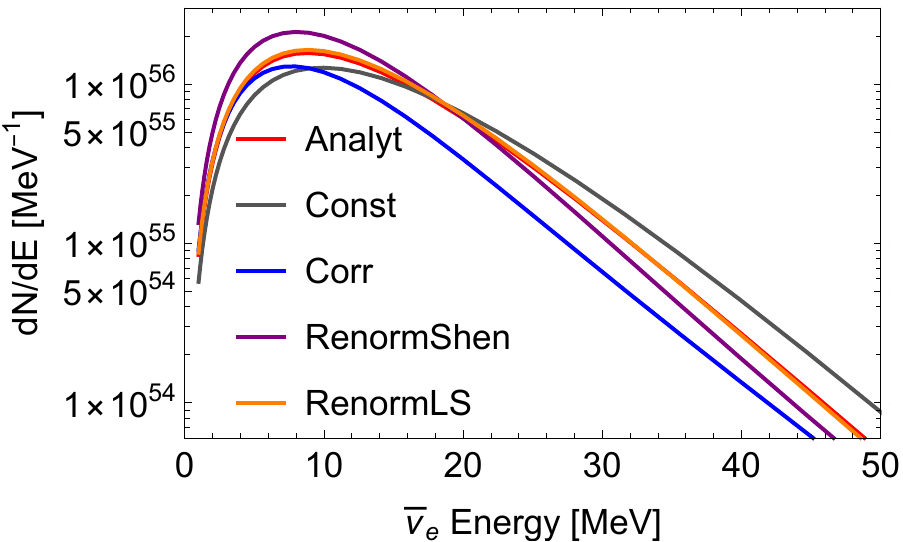}
\caption{Energy spectra for the early + five late-phase neutrino emission estimates. These spectra include the contributions from an ONeMg progenitor and $\sim$8\% failed supernovae for progenitors with initial masses $>40\,M_{\odot}$. The BH neutrino signal in this figure assumes the representative s40 model, but choosing s40s7b2 gives qualitatively similar results. Note that Analyt and RenormLS overlap.}
\label{fig:dNdEfive}
\end{figure}

Specifically, we take Ref.~\cite{2010PhRvL.104y1101H} for the neutrino emission in the ONeMg channel, Burrows and our late-phase strategies for the intermediate masses, and the ``s40'' and ``s40s7b2'' models from Ref.~\cite{2014PhDT.......436H} as two different cases to represent the BH channel. For the successful CCSNe channel, we choose a constant $\alpha=2.3$ to approximate thermal emission and for the failed channel, we use the spectral pinching parameters given by the simulations. In Fig.~\ref{fig:dNdEfive} we show the neutrino energy spectra for each of our five estimates, adopting the ONeMg signal and the s40 model for BH neutrino emission. The spectra using s40s7b2 are qualitatively similar. 

Finally, the DSNB flux is given by the redshift integral over the core-collapse rate:
\begin{equation}
    \frac{d\phi}{dE}=c\int R_{CC}(z)\frac{dN}{dE'}(1+z)\left|\frac{dt}{dz}\right|dz,
\end{equation}
where $E'=E(1+z)$ and $|dt/dz|=H_0(1+z)[\Omega_m(1+z)^3+\Omega_{\Lambda}]^{1/2}$. We integrate up to a maximum redshift of $z=5$, which is sufficient for DSNB contributions (see e.g., Refs.~\cite{2004ando,2017horiuchi,2018moller}). We also assume ``737'' cosmology: $H_0=70\, {\rm km~s^{-1}~Mpc^{-1}},\Omega_M=0.3$, and$~\Omega_{\Lambda}=0.7$ \cite{2006ApJ...636..610R}.
We model the core-collapse rate by:
\begin{equation}
    R_{CC}=\Dot{\rho_*}(z)\frac{\int_{8M_{\odot}}^{100M_{\odot}}\psi(M)dM}{\int_{0.1M_{\odot}}^{100M_{\odot}}M\psi(M)dM},
\end{equation}
where $\Dot{\rho}_*(z)$ is the cosmic star formation rate in units of $M_\odot ~\rm{yr}^{-1}~\rm{Mpc}^{-3}$ from Ref.~\cite{2008yuksel} with parameters from Ref.~\cite{2011horiuchi}. Since the star formation rates are derived assuming a Salpeter IMF, we need a conversion factor to match our assumed IMF; this results in a rescaling factor of 0.55 \cite{2006hopkins}.

\subsection{DSNB event rates}\label{numbers}

We estimate the event rates at SK-Gd where $\sim$10--26$\,{\rm MeV}$ neutrinos are detectable. We calculate the event rate as
\begin{equation}
    R_{\nu}=N_t\int dE\frac{d\phi(E_{\nu})}{dE}\sigma_{\rm IBD}(E_{\nu}),
\end{equation}
where $R_{\nu}$ is the DSNB event rate, $\sigma_{\rm IBD}$ is the inverse beta decay (IBD) cross section as a function of $\overline{\nu}_e$ energy \cite{1999PhRvD..60e3003V}, and $N_t=1.5\times10^{33}$ is the number of IBD targets. 

In Table~\ref{tab:eventresults} we show the estimated DSNB flux and event rates for each of our strategies, where the second and third columns include the s40 BH and the fourth and fifth include the s40s7b2 BH. As to be expected, Const and Corr give the highest and lowest rates, respectively. When integrated, the Analyt and RenormLS methods give very similar rates, while the RenormShen method gives a slightly lower result. The ordering of these rates can be attributed typically to the differences in mean energy, highlighted in Fig.~\ref{fig:fivelate}, except for Corr which is driven also by the the lower liberated energy. The predicted rates vary by a factor $\sim 2$--3. However, excluding Corr, which did not reproduce Bollig well, leads to a min/max range of a factor $\sim 1.5$ which is mostly driven by differences in the neutrino mean energy. 

\begin{table}[b]
\caption{\label{tab:eventresults}
DSNB rate $R_{\nu}$ in ${\rm events~yr^{-1}}$ and integrated flux $\phi$ in ${\rm cm^{-2}~s^{-1}}$ at SK-Gd ($10<E_{\nu}<26\,{\rm MeV}$) with each late-phase strategy. These include the early-phase contribution from the 3D simulations of Ref.~\cite{2019burrows}, an ONeMg progenitor from Ref.~\cite{2010PhRvL.104y1101H}, and a conservative contribution from the failed SNe channel using the neutrino signal from Ref.~\cite{2014PhDT.......436H}. Columns 2 and 3 are the number of events and flux, assuming the neutrino signal from the s40 BH model while columns 4 and 5 assume the s40s7b2 model. These fluxes are well below the current SK upper limits for $E_{\nu}<17.3\,{\rm MeV}$ \cite{Super-Kamiokande:2021jaq}. The Const strategy follows Ref.~\cite{2017horiuchi}, Analyt strategy is based on the work done in Ref.~\cite{2019suwa}, Corr uses the data available from Ref.~\cite{2013nakazato}, and RenormShen/LS uses the data available from Ref.~\cite{2014PhDT.......436H}.}
\begin{ruledtabular}
\begin{tabular}{ccccc}
&\multicolumn{2}{c}{s40 BH}&\multicolumn{2}{c}{s40s7b2 BH}\\
Strategy&$R_{\nu}$ (/yr)&$\phi$ (/cm$^2$/s)&$R_{\nu}$ (/yr)&$\phi$ (/cm$^2$/s)\\
\colrule
\textrm{Const}&2.69&4.57&2.45&4.25\\
\textrm{Analyt}&2.12&3.92&1.88&3.60\\
\textrm{Corr}&1.10&2.14&0.86&1.82\\
\textrm{RenormShen}&1.86&3.73&1.62&3.41\\
\textrm{RenormLS}&2.17&4.04&1.93&3.72\\
\end{tabular}
\end{ruledtabular}
\end{table}

\section{\label{sec:discussion}Discussion and Conclusions}

While recent multidimensional simulations have robust neutrino emission up to the first $\sim1\,{\rm s}$, for the purposes of the DSNB it is necessary to have reasonable estimates for the $\sim$10's of seconds after this since $\gtrsim$ 50\% of the neutrino energy is liberated at these later times. We characterize the PNS cooling phase by estimating the neutrino emission four different ways. These are (i) a constant mean neutrino energy method, (ii) an analytic model for the cooling PNS, (iii) correlations based on the shock revival time and PNS mass, and (iv) rescaled versions of the correlation method.

Based on these four methods, we estimate five DSNB rate predictions (we make two rescaled versions). We include three progenitor populations in our DSNB estimates: collapse of ONeMg cores, collapse of Fe cores to successful SNe, and collapse of Fe cores to black holes. For the ONeMg core channel we take the neutrino emission from Ref.~\cite{2010PhRvL.104y1101H}. For the successfully exploding Fe core channel, we take the neutrino emission computed for the hydrodynamic simulations of Burrows \cite{2019burrows} and add on our five different cooling-phase estimates. For the BH channel we adopt the ``conservative'' estimate of \cite{2018moller}. We ultimately find a factor of $\sim2$--3 difference in the predicted DSNB flux and event rate at SK-Gd, with the constant mean neutrino method (Const) the largest and the correlation method (Corr) the lowest.

It is unsurprising that the constant mean energy strategy overpredicts the DSNB rate: By assuming the mean energy value at the end of the simulations remains constant, it does not model the cooling of PNS evolution. On the low end, we find that compared to other simulations, the 1D simulations of the Supernova Neutrino Database, which drives the correlation method, have lesser liberated and mean energies which results in systematically lower DSNB predictions (although the simulation from Ref.~\cite{2021li} also has similar mean energies out to late times). The renormalized correlation methods (RenormShen and RenormLS) and the analytic solution strategy (Analyt) lie between these two limits. From Fig.~\ref{fig:fivelate}, the five estimates primarily result in mean energy differences while, with the exception of the Corr result, the liberated energies are more similar. An important code comparison study of Ref.~\cite{2018oconnor} showed that near the end of the accretion phase (Fig.~4, $\sim0.5\,{\rm s}$), there is an $\sim\textrm{few}$ MeV difference in mean energies between simulation codes, whereas the luminosities agree well throughout the simulations. These points suggest that, among the neutrino spectral parameters, the uncertainty on the mean energy must be treated carefully. Although it can be seen quantitatively through our cooling-phase estimations, this uncertainty primarily comes from simulation implementation. This is evidenced by the systematic differences between Corr and RenormShen/LS (between the Supernova Neutrino Database and H{\"u}depohl simulations) and by the comparative simulation study \cite{2018oconnor}.

We keep all factors other than the cooling-phase neutrino emission fixed, but these also remain significantly uncertain. For example, there may be significant diversity in the neutrino emission from the BH channel (e.g., s40 vs.~s40s7b2 \cite{2014PhDT.......436H} and different progenitors \cite{2020walk}). Further, simulations of failed SNe and their neutrino emission prove to be strongly EOS dependent, especially regarding the BH formation time \cite{2011oconnor,2013nakazato,2014PhDT.......436H,2019sumiyoshi,2021nakazatotogashi}. In addition, the initial progenitor mass may not be a valid criterion for determining explodability (e.g., \cite{2017horiuchi}); in fact, the $60\,M_{\odot}$ progenitor from the 3D simulations of Ref.~\cite{2019burrows} succeeds in exploding. Although the true failed SNe fraction could be much higher \cite{Horiuchi:2014ska,2018moller,2021kresse,Neustadt:2021jjt}, we keep a more conservative BH fraction so that it minimizes the impact the BH neutrino emission uncertainty has on the DSNB. Some uncertainties that come from implementation details like EOS and dimensionality, are shared between successful and failed cases, but including a smaller BH contribution gives us realistic DSNB results while also establishing the importance of the late-time neutrino emission.

Other factors include the spectral pinching parameter which we have kept fixed to $\alpha=2.3$. Despite being variable at early times, $\alpha$ tends to evolve slowly at longer timescales \cite{2012tamborra,2016NCimR..39....1M}. Its value of 2.3 is largely consistent with the ``best-fit'' procedure of Ref.~\cite{2021kresse}. We do, however, take the appropriate time-integrated $\alpha$ for the BH channels since we extract the data self-consistently from simulations. 

\begin{table}[h]
\caption{\label{tab:otherflavorfluxes}
DSNB integrated flux $\phi$ in ${\rm cm^{-2}~s^{-1}}$ ($10<E_{\nu}<26\,{\rm MeV}$) with each late-phase strategy. These include the early-phase contribution from the 3D simulations of Ref.~\cite{2019burrows}, an ONeMg progenitor from Ref.~\cite{2010PhRvL.104y1101H}, and a conservative contribution from the failed SNe channel using the neutrino signal from Ref.~\cite{2014PhDT.......436H} (s40 model here). Columns 2 and 3 are the fluxes for $\nu_e$ and $\nu_x$, respectively.}
\begin{ruledtabular}
\begin{tabular}{ccc}
Strategy&$\nu_e~\phi$ (/cm$^2$/s)&$\nu_x~\phi$ (/cm$^2$/s)\\
\colrule
\textrm{Const}&3.68&4.13\\
\textrm{Analyt}&2.56&3.60\\
\textrm{Corr}&1.30&1.97\\
\textrm{RenormShen}&2.44&2.84\\
\textrm{RenormLS}&2.75&3.30\\
\end{tabular}
\end{ruledtabular}
\end{table}

Another factor is the overall core-collapse (and/or star formation) rate. Measurements are subject to a number of uncertainties such as disagreement between measured and predicted core-collapse rate \cite{2011horiuchi}, ``invisible'' supernovae \cite{2010lien}, and on the overall normalization \cite{2006hopkins,2011horiuchi}. Additionally, including phenomena like mass transfers and mergers in binary systems can enhance the neutrino signal \cite{2021horiuchi}. Neutrino oscillations like the Mikheyev-Smirnov-Wolfenstein (MSW) effect may also have implications for detection at SK-Gd and other experiments like Hyper-K, DUNE, and JUNO where flavor sensitivity varies \cite{2021tabrizi} and be more impactful in the case of large failed SNe fractions \cite{2021kresse}. We do not include flavor oscillation for simplicity and want to highlight the effects of the cooling-phase estimations. However, we include Table~\ref{tab:otherflavorfluxes} to show the integrated flux with each late phase strategy for $\nu_e$ and $\nu_x$. This highlights that the late-phase strategy chosen is still important for the other neutrino flavors and leads to the same factor of $\sim2$--3 difference and that this conclusion is independent of flavor. Many of these others uncertainties, though, serve to raise or lower the overall rate, not distinguish between different estimations of cooling-phase neutrino emission.

In the future, the most straightforward solution to cooling phase is a large number of long-term ($\sim20\,{\rm s}$), three-dimensional CCSN simulations (see Ref.~\cite{2021bollig} for a recent successful 3D simulation). However, this is almost computationally prohibitive at present. In the meantime, less expensive strategies can be useful. We find that our RenormLS method, where we renormalize the revival-time--final-mass correlations, gives intermediate liberated and mean energies. The correlations themselves (Corr method) and assuming final constant mean energies (Const method) produce systematically too low and high integrated neutrino spectral parameters, respectively. An alternative is to estimate the neutrino luminosity and mean energy with the analytic functional form of Ref.~\cite{2020suwa} and fit these to simulation data (Analyt). However, this method is only preferred if enough simulation data are available past the intense mass accretion phase; it may otherwise lead to unreasonable fits soon after revival time, as in Fig.~\ref{fig:suwasols}. In this context, longer-term two-dimensional simulation sets can be very valuable even if done up to several seconds. 

In conclusion, the factor of $\sim3$ difference in DSNB event rates highlights the relative importance of the late cooling phase and shows that a good understanding of this stage will give more precise DSNB signal estimates, relevant for the upcoming generation of searches.

\begin{acknowledgments}
We thank Hiroki Nagakura and the Princeton supernova simulation group for providing and helping us understand the simulation data used in this work. We also thank Mukul Bhattacharya, Yudai Suwa, and Ken'ichiro Nakazato for helpful discussions. Numerical computations were in part carried out on Cray XC50 at Center for Computational Astrophysics, National Astronomical Observatory of Japan. N.E. is supported by NSF Grant No.~PHY-1914409. The work of S.H. is supported by the U.S.~Department of Energy Office of Science under Award No. DE-SC0020262, NSF Grants No.~AST1908960 and No.~PHY-1914409, and JSPS KAKENHI Grant No. JP22K03630. This study was supported in part by World Premier International Research Center Initiative by the Ministry of Education, Science and Culture of Japan (MEXT), by Grants-in-Aid for Scientific Research of the Japan Society for the Promotion of Science (Grant No.~JP22H01223), the MEXT (Grants No.~JP17H06364, No. JP17H06365, No. JP19H05811, No. JP19K03837, and No. JP20H01905), by the Central Research Institute of Explosive Stellar Phenomena (REISEP) at Fukuoka University, and an associated project (Project No.~207002), and JICFuS as “Program for Promoting Researches on the Supercomputer Fugaku” (Toward a unified view of the universe: from large scale structures to planets, Grant No. JPMXP1020200109). K.S. acknowledges the support by high performance computing resources at Computing Research Center, KEK, Research Center for Nuclear Physics, Osaka University, and Yukawa Institute of Theoretical Physics, Kyoto University.

\end{acknowledgments}

\appendix

\section{Correlations of Other Flavors}\label{corrflavors}

\begin{figure}
\includegraphics[width=\linewidth]{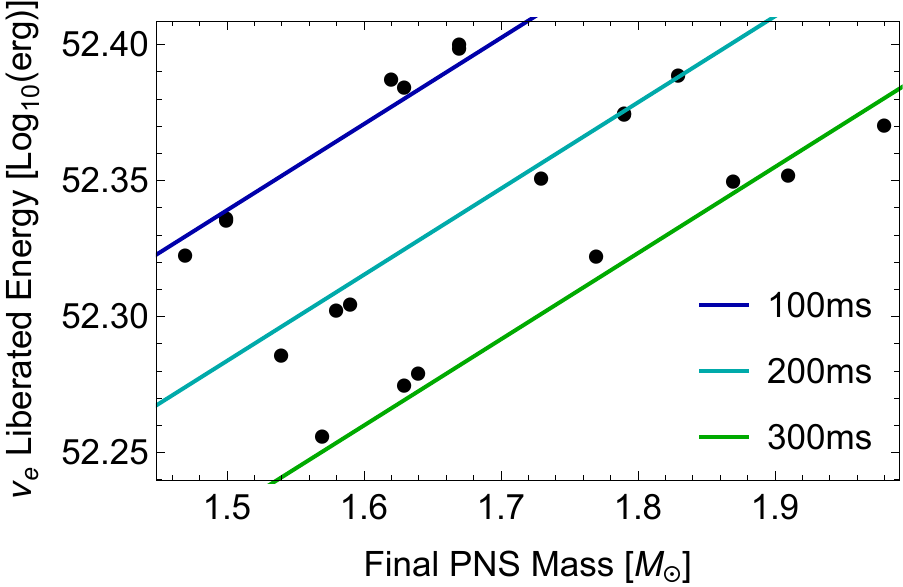}
\includegraphics[width=\linewidth]{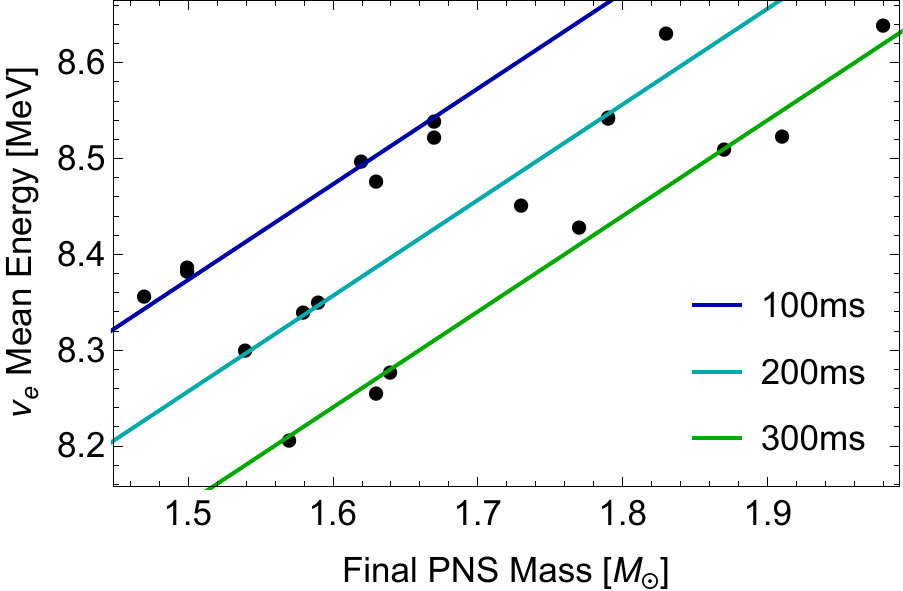}
\caption{Same as Fig.~\ref{fig:ncorrelation} but for $\nu_e$.}
\label{fig:ncorrelationfore}
\end{figure}

\begin{figure}
\includegraphics[width=\linewidth]{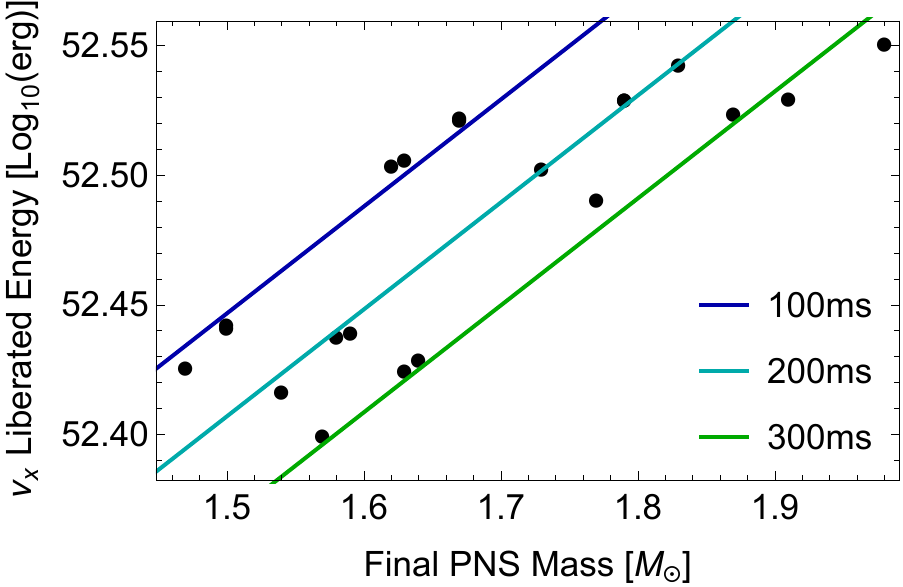}
\includegraphics[width=\linewidth]{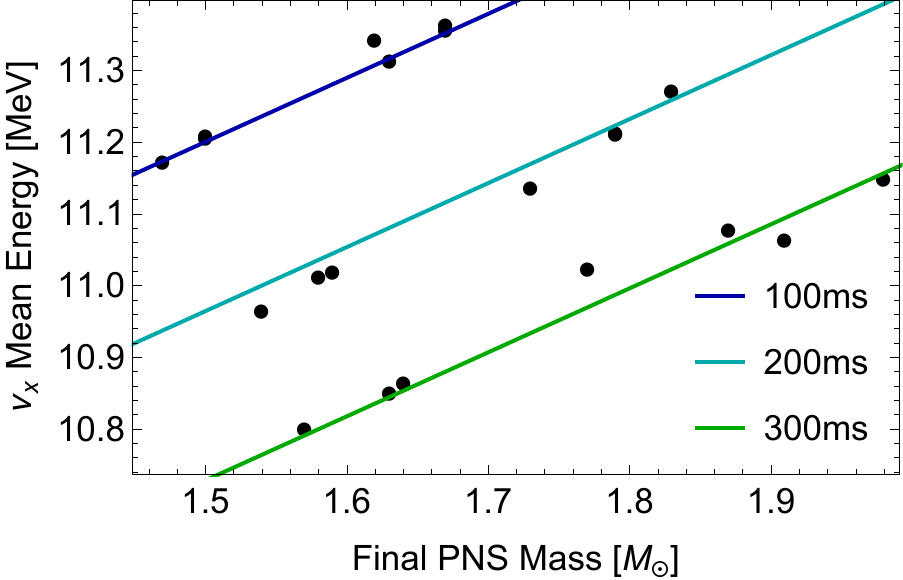}
\caption{Same as Fig.~\ref{fig:ncorrelation} but for $\nu_x$.}
\label{fig:ncorrelationforx}
\end{figure}

\begin{figure}
\includegraphics[width=\linewidth]{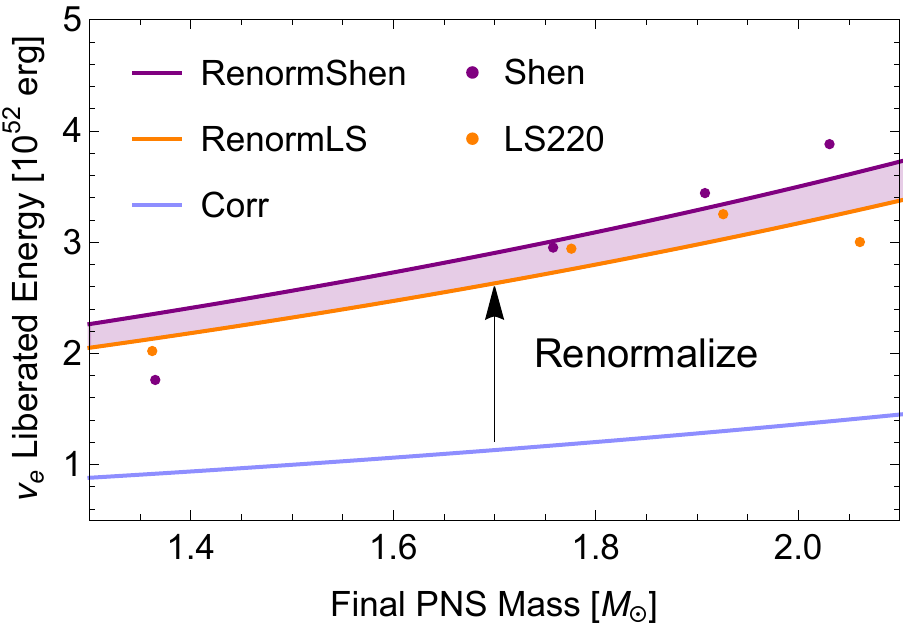}
\includegraphics[width=\linewidth]{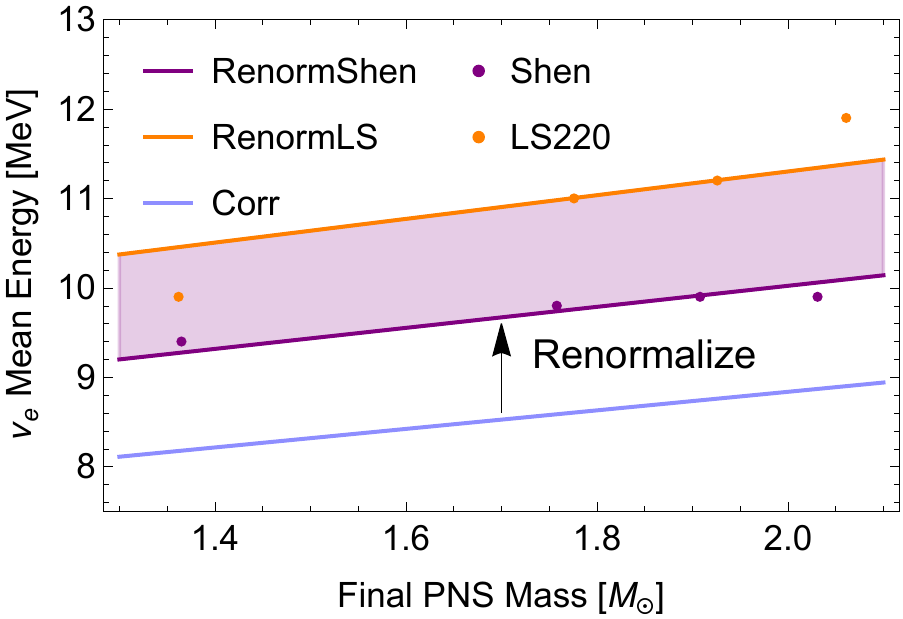}
\caption{Same as Fig.~\ref{fig:hurenorms} but for $\nu_e$.}
\label{fig:hurenormsfore}
\end{figure}

\begin{figure}
\includegraphics[width=\linewidth]{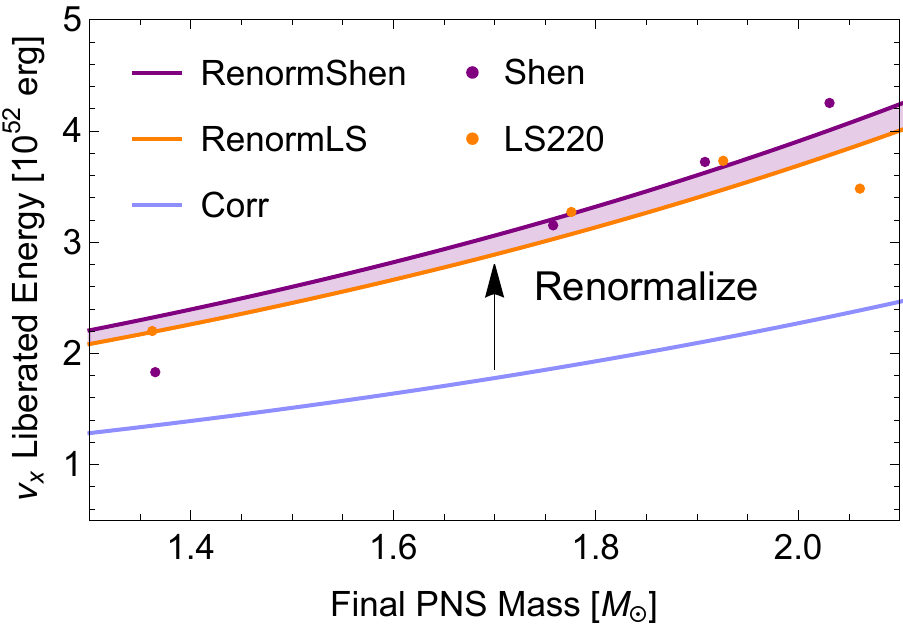}
\includegraphics[width=\linewidth]{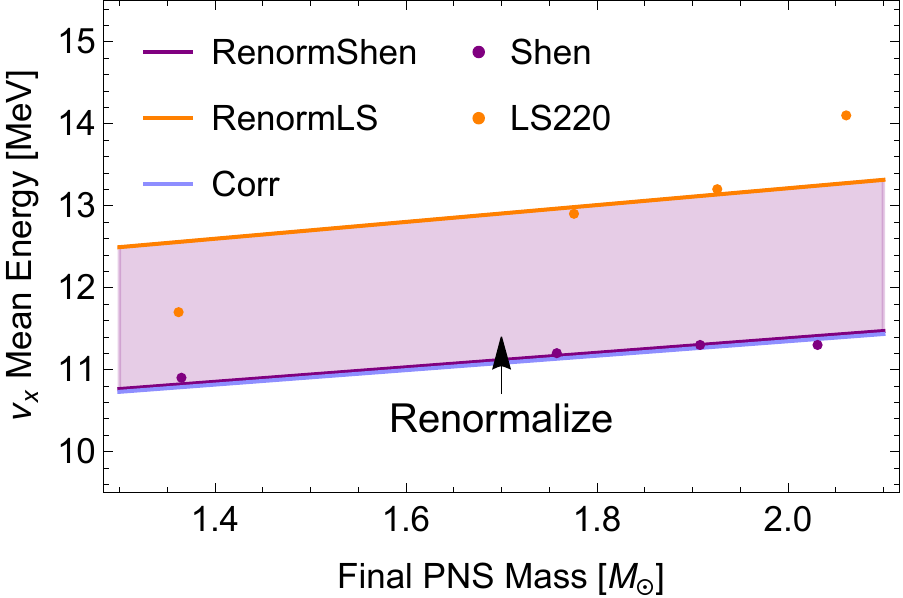}
\caption{Same as Fig.~\ref{fig:hurenorms} but for $\nu_x$. In the bottom panel, the renormalization constant for mean energy is equal to 1 between Corr and RenormShen.}
\label{fig:hurenormsforx}
\end{figure}

In the correlation and renormalization methods (Corr and RenormShen/LS) we find approximately linear correlations between the neutrino spectral parameters, the final baryonic mass of the PNS, and the shock revival time. In Figs.~\ref{fig:ncorrelation} and \ref{fig:hurenorms} we show these for the antielectron neutrino flavor but we find very similar results for $\nu_e$ and $\nu_x$; we see linear trends and renormalized curves that fit through the H{\"u}depohl simulation results well. Here, we show the correlations from the Supernova Neutrino Database for $\nu_e$ and $\nu_x$ in Figs.~\ref{fig:ncorrelationfore} and \ref{fig:ncorrelationforx}. We also show the renormalized curves for $\nu_e$ and $\nu_x$ in Figs.~\ref{fig:hurenormsfore} and \ref{fig:hurenormsforx}.

\bibliography{main}

\end{document}